\documentclass{iopart}
\usepackage{bbm,amssymb,amsbsy,graphicx,subfigure,times}
\usepackage{hyperref}
\usepackage[latin1]{inputenc}
\usepackage{color}

\newcommand{\prl}{{\it Phys. Rev. Lett.} }
\newcommand{\pra}{{\it Phys. Rev. A} }

\newcommand{\qph}{{\it Preprint} quant-ph/}

\newcommand{\gr}[1]{\boldsymbol{#1}}
\newcommand{\sig}{\gr\sigma}
\newcommand{\ket}[1]{|#1\rangle}

\newcommand{\ketbra}[2]{\vert #1 \rangle \! \langle #2 \vert}

\newcommand{\N}{{\cal N}}

\newcommand{\R}{\mathbbm R}
\newcommand{\abs}[1]{\left\vert#1\right\vert}
\renewcommand{\det}{{\rm Det}\,}
\newcommand{\eq}[1]{Eq.~(\ref{#1})}
\newcommand{\ineq}[1]{Ineq.~(\ref{#1})}

\begin{document}

\title[Entanglement sharing of Gaussian states]{\textsf{Continuous variable tangle,
monogamy inequality, and entanglement sharing in Gaussian states of continuous
variable systems}}

\author{\textsf{Gerardo Adesso and Fabrizio Illuminati}}

\address{Dipartimento di Fisica ``E. R. Caianiello'',
Universit\`a degli Studi di Salerno; CNISM and CNR-Coherentia, Gruppo di
Salerno; and INFN Sezione di Napoli-Gruppo Collegato di Salerno; Via
S. Allende, 84081 Baronissi (SA), Italy\\
E-mail: \textcolor[rgb]{0,0,1}{gerardo@sa.infn.it}
  and
\textcolor[rgb]{0,0,1}{illuminati@sa.infn.it}}

\begin{abstract}
For continuous-variable systems, we introduce a measure of
entanglement, the continuous variable tangle ({\em contangle}), with
the purpose of quantifying the distributed (shared) entanglement in
multimode, multipartite Gaussian states. This is achieved by a
proper convex roof extension of the squared logarithmic negativity.
We prove that the contangle satisfies the Coffman-Kundu-Wootters
monogamy inequality in all three--mode Gaussian states, and in all
fully symmetric $N$--mode Gaussian states, for arbitrary $N$. For
three--mode pure states we prove that the residual entanglement is a
genuine tripartite entanglement monotone under Gaussian local
operations and classical communication. We show that pure, symmetric
three--mode Gaussian states allow a promiscuous entanglement
sharing, having both maximum tripartite residual entanglement and
maximum couplewise entanglement between any pair of modes. These
states are thus simultaneous continuous-variable analogs of both the
GHZ and the $W$ states of three qubits: in continuous-variable
systems monogamy does not prevent promiscuity, and the inequivalence
between different classes of maximally entangled states, holding for
systems of three or more qubits, is removed.
\end{abstract}


\clearpage \tableofcontents
\title[Entanglement sharing of Gaussian states]
\noindent \noindent One of the main challenges in fundamental
quantum theory as well as in quantum information and computation
sciences lies in the characterization and quantification of
bipartite entanglement for mixed states, and in the definition and
interpretation of multipartite entanglement both for pure states and
in the presence of mixedness \cite{book1,book2}. More intriguingly,
a quantitative, physically significant, characterization of the
entanglement of states shared by many parties can be attempted: this
approach, introduced in a seminal paper by Coffman, Kundu and
Wootters (CKW) \cite{CKW}, has lead to the discovery of so-called
``monogamy inequalities'', constraining the maximal entanglement
distributed among different internal partitions of a multiparty
system. Such inequalities are uprising as one of the fundamental
guidelines on which proper multipartite entanglement measures have
to be built \cite{sharing}.

While important insights have been gained on these issues in the
context of qubit systems, a less satisfactory understanding has been
achieved until recent times on higher-dimensional systems, as the
structure of entangled states in Hilbert spaces of high
dimensionality exhibits a formidable degree of complexity. However,
and quite remarkably, in infinite-dimensional Hilbert spaces of
continuous-variable systems, important progresses have been obtained
in the understanding of the entanglement properties of a restricted
but fundamental class of states, the so-called Gaussian states
\cite{brareview,eisplenio}. These states, besides being of great
importance both from a theoretical point of view and in practical
applications, share peculiar features that make their structural
properties amenable to accurate and detailed theoretical analysis
\cite{cvbook}.

In this work we address the problem of distributing entanglement
among multiple modes of a continuous variable system. We introduce
the continuous-variable tangle to quantify entanglement sharing in
Gaussian states and we prove that it satisfies the
Coffman-Kundu-Wootters monogamy inequality \cite{CKW}. Nevertheless,
even in the basic instance of three modes, we show that pure,
symmetric Gaussian states, at variance with their discrete-variable
counterparts, allow a promiscuous sharing of quantum correlations,
exhibiting both maximum tripartite residual entanglement and maximum
couplewise entanglement between any pair of modes.

The paper is organized as follows: in Sec.~\ref{secgauss} we review
the basic properties of Gaussian states of continuous variable
systems, and set up notations; in Sec.~\ref{seccontangle} we
address the quantification of entanglement sharing in such states,
introducing a new  entanglement monotone which is shown to
generalize the tangle defined in discrete-variable systems; in
Sec.~\ref{secmono} we apply this measure to prove that all
three--mode Gaussian states and all symmetric multimode Gaussian
states satisfy a monogamy inequality for continuous-variable
entanglement, and that in the specific case of three--mode states
the residual entanglement, emerging from the monogamy inequality, is
a genuine tripartite entanglement monotone; in Sec.~\ref{secstruct}
we exploit this result to investigate the sharing structure of
tripartite entanglement in Gaussian states, unveiling striking
differences with their discrete-variable counterparts; finally, in
Sec.~\ref{secconcl} we summarize our results and outline possible
roadmaps ahead.

\section{Gaussian states: structural properties}\label{secgauss}
In a continuous variable (CV) system consisting of $N$ canonical
bosonic modes, associated to an infinite-dimensional Hilbert space, and
described by the vector $\hat{X}$ of the field quadrature operators,
Gaussian states (such as coherent, squeezed, thermal, and squeezed
thermal states) are those states characterized by first and second
statistical moments of the canonical quadrature operators.
When addressing physical properties, like entanglement, that must
be invariant under local unitary operations, first moments can
be neglected and Gaussian states can then be fully described by the
$2N \times 2N$ real covariance matrix (CM) $\gr{\sigma}$, whose
entries are $\sigma_{ij}=1/2\langle\{\hat{X}_i,\hat{X}_j\}\rangle
-\langle\hat{X}_i\rangle\langle\hat{X}_j\rangle$. This allows, for
Gaussian states, to indicate them indefferently by the density matrix
$\rho$ or by the CM $\gr{\sigma}$. A physical CM
$\gr{\sigma}$ must fulfill the uncertainty relation
\begin{equation}
\label{unciccio}
\gr{\sigma}+i\Omega \geq 0\,,
\end{equation}
with the symplectic form $\Omega=\oplus_{i=1}^{n}\omega$ and
$\omega=\delta_{ij-1}- \delta_{ij+1},\, i,j=1,2.$ Symplectic
operations ({\em i.e.}~belonging to the group $Sp_{(2N,\R)}= \{S\in
SL(2N,\R)\,:\,S^T\Omega S=\Omega\}$) acting by congruence on CMs in
phase space, amount to unitary operations on density matrices in
Hilbert space. In phase space, any $N$--mode Gaussian state can be
written as $\gr{\sigma}= S^T \gr{\nu} S$, with $\gr{\nu}=\,{\rm
diag}\,\{n_1,n_1,n_2,n_2, \ldots, n_N, n_N \}$. The set
$\Sigma=\{n_i\}$ constitutes the symplectic spectrum of
$\gr{\sigma}$ and its elements must fulfill the conditions $n_i\ge
1$, ensuring positivity of the density matrix $\rho$ associated
to $\gr{\sigma}$. The symplectic eigenvalues $n_i$ can be computed
as the eigenvalues of the matrix $|i\Omega\gr{\sigma}|$. The degree
of purity $\mu=\,{\rm Tr}\,\rho^2$ of a Gaussian state with CM
$\gr{\sigma}$ is simply $\mu=1/\sqrt{\,{\rm Det}\,\gr{\sigma}}$.

Concerning the entanglement, positivity of the partially transposed
state $\tilde{\rho}$ (from now on ``$\sim$'' will denote partial
transposition), obtained by transposing the reduced state of only
one of the subsystems, is a necessary and sufficient condition (PPT
criterion) of separability for $(N+1)$--mode Gaussian states of $(1
\times N)$--mode bipartitions \cite{simon00,werner02} and for
$(M+N)$--mode bisymmetric Gaussian states of $(M \times N)$--mode
bipartitions \cite{serafini04}. In phase space, partial
transposition in a $(1 \times N)$--mode bipartition amounts to a
mirror reflection of one quadrature associated to the single--mode
party \cite{simon00}. If $\{\tilde{n}_i\}$ is the symplectic
spectrum of the partially transposed CM $\tilde{\gr{\sigma}}$, then
a $(N+1)$--mode Gaussian state with CM $\gr{\sigma}$ is separable if
and only if $\tilde{n}_i\ge 1$ $\forall\, i$. This implies that a
proper measure of CV entanglement is the {\em logarithmic
negativity} \cite{vidal02}
\begin{equation}
E_{\N} \equiv \ln \|\tilde \rho \|_1 \; , \label{LogNeg}
\end{equation}
where $\| \cdot \|_1$ denotes the trace norm \cite{Fetente}. The
logarithmic negativity $E_{\N}$ is readily computed in terms of the
symplectic spectrum $\tilde{n}_i$ of $\tilde{\gr{\sigma}}$ as
\begin{equation}
E_{\N} = -\sum_{i:\tilde{n}_i<1}\ln \tilde{n}_i \; .
\label{LogNegSymplex}
\end{equation}
Such a measure quantifies the extent to which the PPT condition is
violated. For two--mode symmetric states, the logarithmic negativity
is equivalent to the {\em entanglement of formation} (EoF) $E_{F}$
\cite{efprl}:
\begin{equation}
E_{F}(\rho) \equiv \inf_{\{p_i,|\psi_i \rangle \}} \sum_i p_i
E(|\psi_i \rangle ) \; , \label{EOF}
\end{equation}
where $E(|\psi_i \rangle )$ is the von Neumann entropy (or entropy
of entanglement) of the pure state $|\psi_i \rangle$, and the
infimum is taken over all possible pure states decompositions $\rho
= \sum_{i} p_i |\psi_i \rangle \langle \psi_i |$.

In fact, the logarithmic negativity is positive defined, additive,
monotone under local operations and classical communication (LOCC)
\cite{CapodelFetente}, constitutes an upper bound to the distillable
entanglement in a quantum state $\rho$, and is related to the
entanglement cost under PPT preserving operations \cite{Audenaert}.

\section{The contangle}\label{seccontangle}

Our aim is to analyze the distribution of entanglement between
different (partitions of) modes in CV systems. In Ref.~\cite{CKW}
Coffman, Kundu and Wootters (CKW) proved for system of three qubits,
and conjectured for $N$ qubits (this conjecture has now been proven
by Osborne and Verstraete \cite{Osborne}), that the bipartite
entanglement $E$ (properly quantified) between, say, qubit A
and the remaining two--qubits partition (BC) is never smaller
than the sum of the A$|$B and A$|$C bipartite entanglements
in the reduced states:
\begin{equation}
\label{CKWater}
E^{A|(BC)} \ge E^{A|B} + E^{A|C} \; .
\end{equation}
This statement quantifies the
so-called {\em monogamy} of quantum entanglement \cite{monogam}, in
opposition to the classical correlations, which are not constrained
and can be freely shared.
One would expect a similar inequality to hold for three--mode
Gaussian states, namely
\begin{equation}\label{CKWine}
E^{i|(jk)}- E^{i|j} - E^{i|k} \ge 0 \; ,
\end{equation}
where $E$ is a proper measure of bipartite CV entanglement and the
indexes $\{i,j,k\}$ label the three modes. However, the
demonstration of such a property is plagued by subtle difficulties.
Let us for instance consider the simplest conceivable instance of a
pure three--mode Gaussian state completely invariant under mode
permutations. These pure Gaussian states are named fully symmetric,
and their standard form CM (see Refs. \cite{adescaling,vloock03}),
for any number of modes, is only parametrized by the local mixedness
$a_{loc}=$ $1/\mu_{loc}$, an increasing function of the single--mode
squeezing $r_{loc}$, and $a_{loc} \rightarrow 1^{+}$ when $r_{loc}
\rightarrow 0^{+}$. For these states the inequality (\ref{CKWine})
can be violated for small values of the local squeezing factor,
using either the logarithmic negativity $E_\N$ or the EoF $E_F$
(which is computable in this case, because the two--mode reduced
mixed states of a pure symmetric three--mode Gaussian states are
again symmetric) to quantify the bipartite entanglement. This fact
implies that none of these two measures is the proper candidate for
approaching the task of quantifying entanglement sharing in CV
systems. This situation is reminiscent of the case of qubit systems,
for which the CKW inequality holds using the tangle $\tau$, defined
as the square of the concurrence \cite{Wootters}, but can fail if
one chooses equivalent measures of bipartite entanglement such as
the concurrence itself or the entanglement of formation \cite{CKW}.

It is then necessary to define a proper measure of CV entanglement
that specifically quantifies entanglement sharing according to a
monogamy inequality of the form (\ref{CKWine}). A first important
hint toward this goal comes by observing that, when dealing with
$1\times N$ partitions of fully symmetric multimode pure Gaussian
states together with their $1 \times 1$ reduced partitions, the
desired measure should be a monotonically decreasing function $f$ of
the smallest symplectic eigenvalue $\tilde n_-$ of the corresponding
partially transposed CM $\tilde{\sig}$. This requirement stems from
the fact that $\tilde n_-$ is the only eigenvalue that can be
smaller than $1$ \cite{adescaling}, violating the PPT criterion with
respect to the selected bipartition. Moreover, for a pure symmetric
three--mode Gaussian state, it is necessary to require that the
bipartite entanglements $E^{i|(jk)}$ and $E^{i|j}=E^{i|k}$ be
respectively functions $f(\tilde n_-^{i|(jk)})$ and $f(\tilde
n_-^{i|j})$ of the associated smallest symplectic eigenvalues
$\tilde n_-^{i|(jk)}$ and $\tilde n_-^{i|j}$, in such a way that
they become infinitesimal of the same order in the limit of
vanishing local squeezing, together with their first derivatives:
\begin{equation}
f(\tilde n_-^{i|(jk)})/2f(\tilde n_-^{i|j}) \simeq f'(\tilde
n_-^{i|(jk)})/2f'(\tilde n_-^{i|j}) \rightarrow 1 \; \; \; \; \;
\mbox{for} \; \; \; a_{loc} \rightarrow 1^{+} \; ,
\label{requir}
\end{equation}
where the prime denotes differentiation with respect to the
single--mode mixedness $a_{loc}$. The violation of the sharing
inequality (\ref{CKWine}) exhibited by the logarithmic negativity
can be in fact traced back to the divergence of its first derivative
in the limit of vanishing squeezing. The above condition formalizes
the physical requirement that in a {\it symmetric} state the quantum
correlations should appear smoothly and be distributed uniformly
among all the three modes. One can then see that the unknown
function $f$ exhibiting the desired property is simply the squared
logarithmic negativity \cite{noteconvex}
\begin{equation}\label{funcsq}
f(\tilde n_-)=[-\ln \tilde n_-]^2\,.
\end{equation}
We remind again that for fully symmetric $(N+1)$--mode pure Gaussian
states, the partially transposed CM with respect to any $1 \times N$
bipartition, or with respect to any reduced $1 \times 1$
bipartition, has only one symplectic eigenvalue that can drop below
$1$ \cite{adescaling}; hence the simple form of the logarithmic
negativity (and, equivalently, of its square) in \eq{funcsq}.

 Equipped with this finding,
one can give a formal definition of a bipartite entanglement
monotone that, as we will soon show, can be regarded as the
continuous-variable tangle $E_\tau$. For a generic pure state
$\ket{\psi}$ of a $(1+N)$--mode CV system, we define the square of
the logarithmic negativity:
\begin{equation}
\label{etaupure} E_\tau (\psi) \equiv \ln^2 \| \tilde \rho \|_1 \; ,
\quad \rho = \ketbra\psi\psi \; .
\end{equation}
This is a proper measure of bipartite entanglement, being a convex,
increasing function of the logarithmic negativity $E_\N$, which is
equivalent to the entropy of entanglement for arbitrary pure states.
For any pure multimode Gaussian state $\ket\psi$, with CM $\sig^p$,
of $N+1$ modes assigned in a generic bipartition $1 \times N$,
explicit evaluation gives immediately that
$E_\tau (\psi) \equiv E_\tau (\sig^{p})$ takes the form
\begin{equation}
\label{piupurezzapertutti} E_\tau (\sig^{p}) = \ln^2 \left(1/\mu_1 -
\sqrt{1/\mu_1^2-1}\right) \; ,
\end{equation}
where $\mu_1 = 1/\sqrt{\det\sig_1}$ is the local purity of the reduced
state of mode $1$ with CM $\sig_1$.
Def.~(\ref{etaupure}) is naturally extended
to generic mixed states $\rho$ of $(N+1)$--mode CV systems through
the convex-roof formalism  (see also Ref. \cite{crnega} where a
similar measure, the convex-roof extended negativity, is studied).
Namely, we can introduce the quantity
\begin{equation}\label{etaumix}
E_\tau(\rho) \equiv \inf_{\{p_i,\psi_i\}} \sum_i p_i
E_\tau(\psi_i)\; ,
\end{equation}
where the infimum is taken over all convex decompositions of $\rho$
in terms of pure states $\{\ket{\psi_i}\}$, and if the index $i$ is
continuous, the sum in \eq{etaumix} is replaced by an integral, and
the probabilities $\{p_i\}$ by a probability distribution
$\pi(\psi)$. Let us now recall that, for two qubits, the tangle can
be defined as the convex roof of the squared negativity
\cite{crnega} (the latter being equal to the concurrence
\cite{Wootters} for pure two--qubit states). Here, \eq{etaumix}
states that the convex roof of the squared logarithmic negativity
properly defines the continuous-variable tangle, or, in short, the
{\em contangle} $E_\tau(\rho)$, in which the logarithm takes into
account for the infinite dimensionality of the underlying Hilbert
space.

From now on, we will restrict our attention to Gaussian states. For
any multimode, mixed Gaussian states with CM $\sig$, we will then
denote the contangle by $E_\tau(\sig)$, in analogy with the notation
used for the contangle $E_\tau(\sig^{p})$ of pure Gaussian states in
\eq{piupurezzapertutti}. Any multimode mixed Gaussian state with CM
$\sig$, admits at least one decomposition in terms of pure Gaussian
states $\sig^p$ only. The infimum of the average contangle, taken
over all pure Gaussian state decompositions, defines then the {\it
Gaussian contangle} $G_\tau$:
\begin{equation}
G_\tau(\sig) \equiv \inf_{\{\pi(d\sig^p ), \sig^{p} \}}
\int \pi (d\sig^p) E_\tau (\sig^p) \; .
\label{GaCoRo}
\end{equation}
It follows from the convex roof construction that the Gaussian
contangle $G_\tau(\sig)$ is an upper bound to the true contangle
$E_\tau(\sig)$ (as the latter can be in principle minimized over a
non-Gaussian decomposition):
\begin{equation}
E_\tau(\sig) \leq G_\tau(\sig) \; ,
\label{UpperCut}
\end{equation}
and it can be shown that $G_\tau(\sig)$ is a bipartite entanglement
monotone under Gaussian local operations and classical communication
(GLOCC) \cite{geof,jpa}. In fact, for Gaussian states, the Gaussian
contangle, similarly to the Gaussian EoF \cite{geof}, takes the
simple form
\begin{equation}
G_\tau (\sig) = \inf_{\sig^p \le \sig} E_\tau(\sig^p) \; ,
\label{simple}
\end{equation}
where the infimum runs over all pure Gaussian states with CM $\sig^p
\le \sig$. Let us remark that, if $\sig$ denotes a mixed symmetric
two--mode Gaussian state, then the Gaussian decomposition is the
optimal one \cite{efprl} (it is currently an open question whether
this is true for all Gaussian states \cite{openprob}), and the
optimal pure-state CM $\sig^p$ minimizing $G_\tau(\sig)$ is
characterized by having $\tilde n_-(\tilde {\sig}^p) = \tilde
n_-(\tilde{\sig})$ \cite{geof}. The fact that the smallest
symplectic eigenvalue is the same for both partially transposed CMs
entails that $E_\tau(\sig) = G_\tau(\sig) = [\max\{0,-\ln \tilde
n_-(\sig)\}]^2$. We thus consistently retrieve for the contangle, in
this specific case, the expression previously found for the mixed
symmetric reductions of fully symmetric three--mode pure states,
\eq{funcsq}.

\section{Monogamy inequalities and residual multipartite entanglement}\label{secmono}

\subsection{Monogamy inequality for all three--mode Gaussian states}

We are now in the position to prove the first main result of the
present paper: {\em all} three--mode Gaussian states satisfy the
monogamy inequality (\ref{CKWine}), using the Gaussian contangle
$G_\tau$ (or even the true contangle $E_\tau$ for pure states) to
quantify bipartite entanglement.

We start by considering pure Gaussian states $\sig^p$ of three
modes, each of the three reduced single--mode states being described
respectively by the CMs $\sig_i$, $\sig_j$, $\sig_k$. Due to the
equality of the symplectic spectra across a bipartite cut, following
from the Schmidt decomposition operated at the CM level
\cite{holewer}, any one of the two--mode reduced CMs $\sig_{ij}$,
$\sig_{ik}$, $\sig_{jk}$, will have smallest symplectic eigenvalue
of the associated partially transposed CM equal to 1, and will thus
represent a mixed state of partial minimum uncertainty
\cite{extremal,extremal2}. Given the three modes $i,j,k$, any such
two--mode reduction, say that of modes $i$ and $j$, is completely
specified, up to local single--mode unitary operations, by the two
local purities $\mu_{i}=(\det\sig_{i})^{-1/2}$ and
$\mu_{j}=(\det\sig_{j})^{-1/2}$, and by the global two--mode purity
equal to the local purity of mode $k$:
$\mu_{ij}=(\det\sig_{ij})^{-1/2}=(\det\sig_{k})^{-1/2}=\mu_k$.
Moreover, because the contangle between mode $i$ and modes $(jk)$ is
a function of $\mu_i$ alone, see \eq{piupurezzapertutti}, all the
entanglement properties of three--mode pure Gaussian states are
completely determined by the three local purities $\mu_i$, $\mu_j$,
and $\mu_k$, or by the associated local mixednesses  $a_i \equiv
1/\mu_i$, $a_j \equiv 1/\mu_j$, and $a_k \equiv 1/\mu_k$. The local
mixednesses $a_i$, $a_j$, and $a_k$ have then to vary constrained by
the triangle inequality
\begin{equation}
\label{triangle}
|a_j - a_k| + 1 \; \le \; a_i \; \le \; a_j + a_k - 1 \; ,
\end{equation}
in order for $\sig^p$ to be a physical state. This is a
straightforward consequence of the uncertainty relation
Ineq.~(\ref{unciccio}) applied to the reduced states of any two
modes (see \cite{3modi,jpa} for further details). Notice that
\ineq{triangle} is a stronger requirement than the general
Araki-Lieb inequality holding for the Von Neumann entropies of each
single--mode reduced state \cite{3modi}. For ease of notation, let
us rename the mode indices: $\{i,j,k\} \equiv \{1,2,3\}$. Without
loss of generality, we can assume $a_1 > 1$ (if $a_1=1$ the first
mode is not correlated with the other two and all terms in
\ineq{CKWine} are trivially zero). Moreover, we can restrict to the
case of $\sig_{12}$ and $\sig_{13}$ being both entangled. In fact,
if {\em e.g.}~$\sig_{13}$ denotes a separable state, then
$E_\tau(\sig_{12}) \le E_\tau^{1|(23)}(\sig^p)$ because tracing out
mode $3$ is a LOCC (see Ref. \cite{vidal02}), and thus the sharing
inequality is automatically satisfied.

We will now prove \ineq{CKWine} in general by using the Gaussian
contangle, as it will imply the inequality for the true contangle;
in fact $G_\tau^{1|(23)}(\sig^p) = E_\tau^{1|(23)}(\sig^p)$ but
$G_\tau^{1|l}(\sig) \ge E_\tau^{1|l}(\sig),\, l=2,3$. Our strategy
will be to show that, at fixed $a_1$, {\em i.e.}~at fixed
entanglement between mode $1$ and the remaining modes:
\begin{equation}
E_\tau^{1|(23)} = \ln^2(a_1-\sqrt{a_1^2-1}) \; ,
\end{equation}
the maximum value of the sum of the $1|2$ and $1|3$ bipartite
entanglements can never exceed $E_\tau^{1|(23)}$. Namely,
$\max_{s_1,d_1} Q  \le \ln^2(a_1-\sqrt{a_1^2-1})$, where
\begin{equation}
Q \equiv G_\tau(\sig_{12}) + G_\tau(\sig_{13}) \; ,
\end{equation}
and the variables $a_2$ and $a_3$ have been replaced by
the variables $s_1 = (a_2+a_3)/2$ and $d_1 = (a_2-a_3)/2$.
The latter are constrained to vary in the region
\begin{equation}\label{region}
s_1 \ge \frac{a_1+1}{2}\,,\quad\abs{d_1} \le \frac{a_1^2-1}{4s_1}
\,,
\end{equation}
defined by the triangle inequality (\ref{triangle}) and
by the condition of the reduced two--mode bipartitions being
entangled \cite{jpa}. We recall now that each $\sig_{1l}$, $l=2,3$,
is a two--mode state of partial minimum uncertainty. For this class
of states the Gaussian measures of entanglement, including $G_\tau$,
can all be determined and have been computed explicitely \cite{jpa}.
Skipping straightforward but tedious calculational details, and
omitting from now on the subscript $1$, we have that
\begin{eqnarray}
Q & = & \ln^2 [m(a,s,d) - \sqrt{m^2(a,s,d)-1}] \nonumber \\
& + & \ln^2 [m(a,s,-d) - \sqrt{m^2(a,s,d)-1}] \; ,
\label{Quisizione}
\end{eqnarray}
where $m = m_-$ if $D \le 0$, $m = m_+$ otherwise, and
$$m_- \equiv |k_-|/[(s-d)^2-1] \; ,$$
$$m_+ \equiv \{2\,[2 a^2 (1+2 s^2 + 2 d^2) - (4 s^2 - 1)(4 d^2 - 1) -a^4 -
\sqrt{\delta}]\}^{1/2}/[4(s-d)] \; ,$$
$$D = 2 (s - d) - \sqrt{2\left[k_-^2 + 2 k_++|k_-| (k_-^2 + 8
k_+)^{1/2}\right]/k_+} \; ,$$
$$k_\pm = a^2 \pm (s+d)^2 \; ,$$
$$\delta = (a - 2 d - 1) (a - 2 d + 1) (a + 2 d - 1) (a + 2 d + 1) (a - 2 s -
1) (a - 2 s + 1) (a + 2 s - 1) (a + 2 s + 1) \; .$$

Studying the derivative of $m_\mp$ with respect to $s$, it is
analytically proven that, in the whole space of parameters
$\{a,s,d\}$ given by \eq{region}, both $m_-$ and $m_+$ are
monotonically decreasing functions of $s$. The quantity $Q$ is then
maximized over $s$ for the limiting value $s = s^{\min} \equiv (a +
1)/2$. This value corresponds to three--mode pure states in which
the reduced partition $2|3$ is always separable, as it is intuitive
because the bipartite entanglement is maximally concentrated in the
$1|2$ and $1|3$ partitions. With the position $s = s^{\min}$, $D$
can be easily shown to be always negative, so that, for both reduced
CMs $\sig_{12}$ and $\sig_{13}$, the Gaussian contangle is defined
in terms of $m_-$. The latter, in turn, acquires the simple form
$$m_-(a,s^{\min},d) = \frac{1 + 3a + 2d}{3 + a - 2 d}\,.$$
Consequently, the quantity $Q$ is immediately seen to be an even and
convex function of $d$, which entails that it is globally maximized
at the boundary $|d| = d^{\max} \equiv (a-1)/2$. It turns out that
\begin{equation}\label{finalised}
Q[a,s=s^{\min},d=\pm d^{\max}]=\ln^2(a-\sqrt{a^2-1})\,,
\end{equation}
which implies that in this case the sharing inequality
(\ref{CKWine}) is saturated and the genuine tripartite entanglement
is exactly zero. In fact this case yields states with $a_2=a_1$ and
$a_3=1$ (if $d=d^{\max}$), or $a_3=a_1$ and $a_2=1$ (if
$d=-d^{\max}$), {\em i.e.}~tensor products of a two--mode squeezed
state and a single--mode uncorrelated vacuum. Being the above
quantity \eq{finalised} the global maximum of $Q$, \ineq{CKWine}
holds true for any $Q$, that is for {\em any} pure three--mode
Gaussian state, choosing either the Gaussian contangle $G_\tau$ or
the true contangle $E_\tau$ as measures of bipartite entanglement.

By convex roof construction, the above proof immediately extends
to all mixed three--mode Gaussian states $\sig$, with the bipartite
entanglement measured by $G_\tau$.
Let $\{\pi(d\sig^p_{m}), \sig^{p}_{m}\}$ be the ensemble of pure
Gaussian states minimizing the Gaussian convex roof in
Eq.~(\ref{GaCoRo}); then, we have
\begin{eqnarray}
G_\tau^{i|(jk)}(\sig) & = &
\int \pi (d\sig^p_{m}) G_\tau^{i|(jk)}(\sig^p_{m}) \nonumber \\
& \ge & \int \pi (d\sig^p_{m})
[G_\tau^{i|j}(\sig^p_{m}) + G_\tau^{i|k}(\sig^p_{m})] \nonumber \\
& \ge & G_\tau^{i|j}(\sig) + G_\tau^{i|k}(\sig) \; ,
\label{GlandePlova}
\end{eqnarray}
where we have exploited the fact that the Gaussian contangle
is convex by construction. This concludes the proof of the
monogamy inequality~(\ref{CKWine}) for all three--mode
Gaussian states. $\hfill \blacksquare$

\subsection{Residual tripartite entanglement and monotonicity}

The sharing constraint (\ref{CKWine}) leads naturally to the
definition of the {\em residual contangle} as a quantifier of
genuine tripartite entanglement ({\em arravogliament}). This is in
complete analogy with the case of qubit systems, except that, at
variance with the three-qubit case (where the residual tangle of
pure states is invariant under qubit permutations), for CV systems
of three modes the residual contangle is partition-dependent
according to the choice of the reference mode (but for the fully
symmetric case). Then, the {\em bona fide} quantification of
tripartite entanglement is provided by the {\em minimum} residual
contangle:
\begin{equation}
\label{etaumin}
E_\tau^{i|j|k}\equiv\min_{(i,j,k)} \left[
E_\tau^{i|(jk)}-E_\tau^{i|j}-E_\tau^{i|k}\right] \; ,
\end{equation}
where $(i,j,k)$ denotes all the possible permutations of the three
mode indexes. This definition ensures that $E_\tau^{i|j|k}$ is invariant
under mode permutations and is thus a genuine three-way property of
any three--mode Gaussian state. One can verify that
$$E_\tau^{i|(jk)}-E_\tau^{i|k}-(E_\tau^{j|(ik)}-E_\tau^{j|k}) \ge 0$$
if and only if $a_i \ge a_j$, and therefore the absolute
minimum in \eq{etaumin} is attained by the decomposition realized
with respect to the reference mode $i$ of smallest local mixedness
 $a_i$, i.e. of largest local purity $\mu_{i}$.

A crucial requirement for $E_\tau^{i|j|k}$ to be a proper measure of
tripartite entanglement is that it be nonincreasing under LOCC. The
monotonicity of the residual tangle was proven for three--qubit pure
states in Ref. \cite{wstates}. In the CV setting, we will now prove
that for pure three--mode Gaussian states the residual Gaussian
contangle $G_\tau^{i|j|k}$, defined in analogy with \eq{etaumin}, is
an {\em entanglement monotone} under tripartite GLOCC, and
specifically that it is nonincreasing even for probabilistic
operations, which is a stronger property than being only monotone on
average \cite{geof}. We thus want to prove that
$$G_\tau^{i|j|k}(G_p(\sig^p)) \le G_\tau^{i|j|k}(\sig^p)$$, where
$G_p$ is a pure GLOCC mapping pure Gaussian states $\sig^{p}$ into
pure Gaussian states \cite{giedjens,Scheel}. Every GLOCC protocol
can be realized through a local operation on one party only. Assume
that the minimum in \eq{etaumin} is realized for the reference mode
$i$; the output of a pure GLOCC $G_p$ acting on mode $i$ yields a
pure-state CM with $a_i' \le a_i$, while $a_{j}$ and $a_{k}$ remain
unchanged \cite{giedjens}. Then, the monotonicity of the residual
Gaussian contangle $G_\tau^{i|j|k}$ under GLOCC is equivalent to
proving that $G_\tau^{i|j|k} =
G_\tau^{i|(jk)}-G_\tau^{i|j}-G_\tau^{i|k}$  is a monotonically
increasing function of $a_i$ for pure Gaussian states. One can
indeed show that the first derivative of $G_\tau^{i|j|k}$ with
respect to $a_i$, under the further constraint $a_i \le a_{j,k}$, is
globally minimized for $a_i=a_j=a_k \equiv a_{loc}$, {\em i.e.}~for
a fully symmetric state. It is easy to verify that this minimum is
always positive for any $a_{loc} >1$, because in fully symmetric
states the residual contangle is an increasing function of
$a_{loc}$. Therefore the monotonicity of $G_\tau^{i|j|k}$ under
GLOCC for all pure three--mode Gaussian states is finally proven.
$\hfill \blacksquare$

\subsection{Monogamy inequality for $N$--mode symmetric Gaussian states}

We next want to investigate whether the monogamy
inequality (\ref{CKWine}) can be generalized to Gaussian
states with an arbitrary number $N+1$ of modes, namely whether
\begin{equation}\label{monoN}
E^{i|(j_1 , \ldots , j_N)} - \sum_{l=1}^{N} E^{i|j_l} \geq 0 \; .
\end{equation}
Establishing this result in general is a highly nontrivial
task, but we can readily prove it for all {\it symmetric} multimode
Gaussian states. As usual, due to the convexity of $G_\tau$, it will
suffice to prove it for pure states, for which the
Gaussian contangle coincides with the true contangle in every
bipartition. For any $N$ and for $a_{loc}> 1$ (for $a_{loc}=1$ we
have a product state),
\begin{equation}
E_\tau^{i|(j_1 , \ldots , j_N)} = \ln^2(a_{loc}-\sqrt{a_{loc}^2-1})
\end{equation}
is independent of $N$, while the total two--mode
contangle
\begin{eqnarray}
N E_\tau^{i|j_l} & = & \frac{N}{4} \ln^2 \bigg\{ \Big[ a_{loc}^2 (N+1) - 1  \nonumber \\
& - &  \sqrt{(a_{loc}^2-1)(a_{loc}^2(N+1)^2-(N-1)^2)} \; \; \Big] /
N \bigg\}
\end{eqnarray}
is a monotonically decreasing function of the integer $N$ at fixed
$a_{loc}$.  Because the sharing inequality trivially holds for
$N=1$, it is inductively proven for any $N$. $\hfill \blacksquare$

\begin{figure}[t!]
\centering{
\includegraphics[width=8cm]{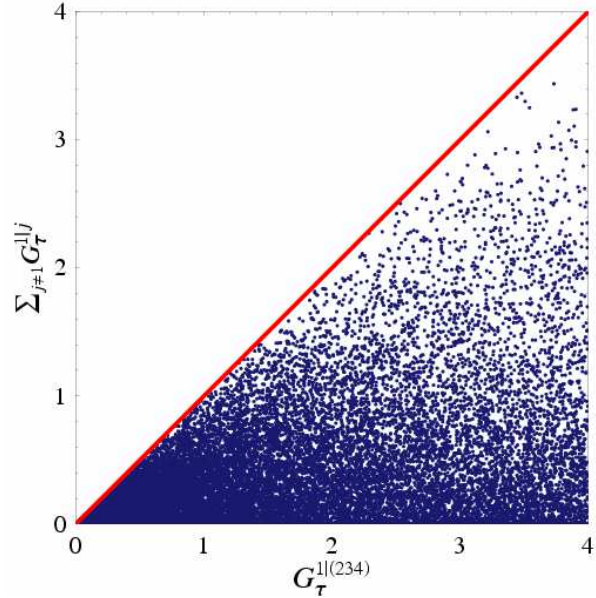}
\caption{Behaviour of the Gaussian contangle $G_\tau^{1|(234)}$ between
mode $1$ and all the other remaining three modes (horizontal axis),
plotted versus the total bipartite Gaussian contangle $\sum_{j=2}^{4} G_\tau^{1|j}$
(vertical axis). The two quantities have been evaluated numerically in
60000, randomly generated, four--mode Gaussian states.
States saturating the monogamy inequality Eq. (\ref{monoN}) fall on the
solid red line. No physical state has been found lying in the upper-left half
of the plane, above the red line. This numerical result supports the conjecture
that the Gaussian contangle is monogamous, and satisfies Eq. (\ref{monoN})
in all multimode Gaussian states. It is important to notice that, for increasing
entanglement between one mode and the others, the states tend to distribute farther
away from the 45-degree line boundary, hinting at the presence of genuine multipartite
entanglement shared between all the modes.}
\label{ficarra}}
\end{figure}

This result, together with extensive numerical evidence obtained for
randomly generated non-symmetric $4$--mode Gaussian states (see
Fig.~\ref{ficarra}), strongly supports the conjecture that the
monogamy inequality be true for {\em all} multimode Gaussian state,
using the (Gaussian) contangle as a measure of bipartite
entanglement . It is currently under way to fully prove the
conjecture analytically.

\section{Structure of entanglement sharing: the CV GHZ/$W$ states}\label{secstruct}

We are now in the position to analyze the sharing structure of CV
entanglement by taking the residual contangle as a measure of
tripartite entanglement, in analogy with the study of three-qubit
entanglement \cite{wstates}. Namely, we pose the problem of
identifying the three--mode analogues of the two inequivalent
classes of fully inseparable three--qubit states, the GHZ state
\cite{ghzs} $\ket{\psi_{\rm GHZ}} = (1/\sqrt{2}) \left[\ket{000} +
\ket{111}\right]$, and the $W$ state \cite{wstates} $\ket{\psi_{W}}
= (1/\sqrt{3}) \left[\ket{001} + \ket{010} + \ket{100}\right]$.
These states are pure and fully symmetric. On the one hand, the GHZ
state possesses maximal tripartite entanglement, quantified by the
residual tangle \cite{CKW,wstates}, without any two--qubit
entanglement. On the other hand, the $W$ state contains the maximal
two-party entanglement between any pair of qubits \cite{wstates} and
its tripartite residual tangle is consequently zero.

Surprisingly enough, in symmetric three--mode Gaussian states, if one aims at
maximizing (at given single--mode squeezing $r_{loc}$ or single--mode
mixedness $a_{loc}$) either the two--mode contangle $E_\tau^{i|l}$
in any reduced state ({\it i.e.}~aiming at finding the CV analogue
of the $W$ state), or the genuine tripartite contangle
({\it i.e.}~aiming at defining the CV analogue of the GHZ state),
one finds the same, unique family of pure symmetric three--mode squeezed
states. These states, previously named ``GHZ-type'' states \cite{vloock03},
can be defined for generic $N$--mode systems, and constitute an ideal
test-ground for the study of multimode CV entanglement \cite{adescaling,serafini04}.
The peculiar nature of entanglement sharing in this class of CV GHZ/$W$
states is further confirmed by the following observation. If one
requires maximization of the $1 \times 2$ bipartite contangle
$E_\tau^{i|(jk)}$ under the constraint of separability of all
two--mode reductions, one finds a class of symmetric mixed states
whose tripartite residual contangle is strictly smaller than the one
of the GHZ/$W$ states, at fixed local squeezing. Therefore, in
symmetric three--mode Gaussian states, when there is no two--mode
entanglement, the three-party one is not enhanced, but frustrated.
Instead, if a mode is maximally entangled with another, it can also
achieve maximal quantum correlations in a three-party relation.

These results, unveiling a major difference between
discrete-variable and CV systems, establish the {\em promiscuous}
nature of CV entanglement sharing in symmetric Gaussian states.
Being associated to degrees of freedom with continuous spectra,
states of CV systems need not saturate the CKW inequality to achieve
maximum couplewise correlations. In fact, without violating the
monogamy inequality (\ref{CKWine}), pure symmetric three--mode
Gaussian states are maximally three-way entangled and, at the same
time, maximally robust against the loss of one of the modes due, for
instance, to decoherence.

Furthermore, the residual contangle \eq{etaumin} in GHZ/$W$ states
acquires a clear operative interpretation in terms of the optimal
fidelity in a three-party CV teleportation network
\cite{telepoppate}. This finding readily provides an experimental
test, in terms of success of teleportation-network experiments
\cite{bra00,naturusawa}, to observe the promiscuous distribution of
CV entanglement in symmetric, three--mode Gaussian states.

\section{Concluding remarks}\label{secconcl}

It is a central trait of quantum information theory that there exist
limitations to the free sharing of quantum correlations among
multiple parties \cite{sharing}. This aspect can be quantified in
terms of {\em monogamy constraints}, as first introduced by Coffman,
Kundu and Wootters for states of three qubits. In this paper we have
generalized these monogamy constraints to infinite-dimensional
systems. This extension required the definition of a proper
entanglement monotone, able to capture the trade-off between the
couplewise entanglement and the genuine tripartite and, in general,
multipartite entanglement in multimode Gaussian states. We proved
analytically that the continuous-variable entanglement is monogamous
in all three--mode and in all symmetric multimode Gaussian states,
and have numerically convincing evidence that this holds true in
{\it all} $N$--mode Gaussian states as well.

Very remarkably, in the case of pure states of three modes, the
residual entanglement emerging from the monogamy inequality turns
out to be a tripartite entanglement monotone under Gaussian LOCC,
representing the first {\em bona fide} measure of genuine
multipartite entanglement for CV systems. This measure has been
applied to investigate the concrete structure of the distributed
entanglement in three--mode Gaussian states, leading to the
discovery that there exists a special class of states (pure,
symmetric, three--mode squeezed states) which simultaneously
maximize the genuine tripartite entanglement {\it and} the bipartite
entanglement in the reduced states of any pair of modes. This
property, which has no counterpart in finite-dimensional systems,
has been labeled by us {\em promiscuous sharing} of CV entanglement.

The collection of results presented here is of basic importance for
the understanding of quantum correlations among multiple parties in
systems with infinitely many degrees of freedom. Several hints
emerging from our results are worth being investigated. Among them,
the analysis of the effect of decoherence on states with promiscuous
entanglement sharing, to study the actual robustness of this
correlation structure in the presence of noise and mixedness. From a
more fundamental point of view, the extension of the present
techniques to define a more general measure of multipartite
entanglement for all multimode, Gaussian and non-Gaussian states is the
most evident, although very challenging, task to aim at.
Another important follow-up should concern the experimental
implications of our findings, ranging from the verification of
the promiscuity in terms of quantum communication experiments
\cite{telepoppate}, to the preparation of special classes of
multiparty entangled Gaussian states and the implementation of
these resources for novel protocols of quantum information with
continuous variables. All these issues are being currently
and actively studied \cite{3modi}.

In a wider perspective, we wish to mention that monogamy
inequalities should play a fundamental role in the understanding of
entanglement and other structural properties of a wide class of
complex quantum systems both in discrete and continuous variables
\cite{Osborne, fazio, OsborneNielsen, Roscilde, Plenio}. At this
stage, in the continuous-variable setting, the most interesting and
urgent open problems are perhaps the extension of the CKW-type
sharing inequalities to non-Gaussian states and to systems in which
the reference party is constituted by more than one mode. The latter
problem is the CV analogue of extending the monogamy inequalities to
systems in which the reference party is constituted by more than one
qubit \cite{Osborne}. Finally, another very intriguing open issue is
the generalization to systems of arbitrary finite dimension $D$
interpolating between the case $D=2$ and the infinite-dimensional
instance \cite{sharing}. In all these problems the challenge is to
define easily computable measures of entanglement that allow
concrete quantifications and bounds on the sharing structure and
distribution of quantum correlations.

\addcontentsline{toc}{section}{Acknowledgments}
\section*{Acknowledgements} \noindent We thank O. Kr\"uger, A. Serafini, and W. K.
Wootters for stimulating discussions and helpful comments. Financial
support from CNR-Coherentia, INFN, and MIUR is acknowledged.


\begin{thebibliography}{99}
\bibitem{noteconvex}
Notice that an infinite number of functions satisfying \eq{requir}
can be obtained by expanding $f(\tilde n_-)$ around $\tilde n_- = 1$
at any even order. However, they are all convex functions of $f$. If
the inequality (\ref{CKWine}) holds for $f$, it will hold as well
for any monotonically increasing, convex function of $f$, such as
the logarithmic negativity raised to any even power $k \ge 2$, but
not for $k=1$.
\end{thebibliography}
\end{document}